\documentclass[9pt,twocolumn]{article} 
\usepackage{times}
\usepackage{graphicx}
\usepackage{amssymb}
\usepackage{mathtools}
\usepackage{commath}
\usepackage{url,hyperref}
\usepackage{cite}
\usepackage{arxivPaper}

\hypersetup{
	urlcolor=blue,
	colorlinks=true,
	allcolors=blue,
}

\begin{document}

\title{Revisited theory of selective reflection from a dilute Fabry-Perot interferometer}

\author{Davit Khachatryan\\
\normalsize{\it{Institute for Physical Research, NAS of Armenia, Ashtarak-2, 0203 Armenia}}\\
\normalsize{\it{davit.khachatryan@email.com}}\\
\normalsize{\it{\today}}
}

\maketitle

\begin{abstract}
In this paper we revisited the theory of selective reflection from a dilute vapor cell. A self-consistent theory was developed for reflection spectrum for Fabry-Perot interferometer. Formulas for single and multiple reflections were obtained. We also obtained the effective refractive index for a single selective reflection. The results of the paper are in a good agreement with existing experimental results. 
\end{abstract}

\section{Introduction}
\label{sec::intro}
Reflection of radiation from the boundary between a dielectric and atomic vapor, when laser field is detuned in the vicinity of atomic transition frequencies, is termed as selective reflection (SR) \cite{woerdman1975, Schuurmans1976}. SR has many applications, such as locking a diode laser frequency to atomic resonance lines \cite{gazazyan2007, li1998, muller1998}, retrieval of group refractive index \cite{Papoyan2017},  marking atomic transition resonance lines \cite{Klinger2017JETP, Sargsyan2016},  study of the van-der-Waals interaction of atoms with a dielectric surface \cite{failache2003, fichet2007, Sargsyan2017OpLe}, determination of the homogeneous width and the shift of resonance lines \cite{vuletic1993, Wang1997, papoyan1998Izvest, Sargsyan2017OpLe} and cross-sections of resonant collisions \cite{papoyan1998OpSp}, study of coherent and magneto-optical processes \cite{gross1997, weis1992, papageorgiou1994, Sargsyan2017JOSAB, Klinger2017Euro, Sargsyan2018}.

In the recent decades the theory of SR was investigated by many authors \cite{Schuurmans1976, ducloy1991, Vartanyan2001, guo1994, guo1996, Vartanyan1995, Dutier2003}. Schuurmans in his paper \cite{Schuurmans1976} developed a theory, where he obtained the spectral narrowing of a SR signal. In \cite{ducloy1991} a theory of frequency modulated SR was developed, that allows one to obtain Doppler free spectral lines. The problem of different frequency shifts existing in a SR signal was discussed by \cite{Vartanyan2001, guo1994, guo1996} (e. g. caused by local-field correction, atom-wall
interaction, non-exponential attenuation of the field in the vapor). In \cite{Vartanyan1995, Dutier2003} theories were developed for the dilute thin vapor cells.

The sub-Doppler reflection spectrum in a SR signal is due to atom-wall collisions. After a collision atoms leave the wall in the ground state. This creates a spatial transient region, where the polarization has dependence on the spatial coordinate. The existence of this transient region was experimentally demonstrated by several authors \cite{Burgmans1977, bordo1997} by using evanescent wave fluorescence spectrum from the atoms that are near the dielectric wall.

In \cite{Papoyan2004} it was shown that by changing the length of the cell's highly parallel window one can change the SR signal shape, because of the Fabry-Perot interferometer effect. Also, one can change the SR signal shape by changing the length of the thin (nanometric) vapor cell \cite{Sargsyan2016}. In this paper we demonstrated that the SR line shape can be changed also for a thicker (cm order) cell.

This paper is an extension of the theory presented in \cite{Papoyan2017}. We developed a self-consistent theory by using the density matrix formalism and Maxwell equations. In the calculations we used Laplace transformation that let us obtain formulas for single and multiple selective reflection. Also, the effective complex refractive index for a single selective reflection was obtained. We assume that atom-wall collisions are diffusive, that is, after a collision all atoms lose their polarization \cite{Dutier2003, Vartanyan1995}. Since light is directed normally to the cell boundary, we consider the one-dimensional problem. 

In section \ref{sec::Basic} we will show some classical formulas for reflection from a Fabry-Perot interferometer and for the laser field inside the cell by taking into account the steady-state solution. Then, in section \ref{sec::SpDisp} we will take into account the transient behavior of polarization of a dilute vapor cell. We will obtain new formulas for single and multiple reflection spectra and discuss the considered approximations. Additionally, we will compare the new formulas with the classical ones. In section \ref{sec::RefractiveIndex} we will present the effective refractive index and discuss its properties. Finally, in section \ref{sec::Selective} we will show the results obtained from our formulas and compare them with existing experimental results.

\section{Basic concepts}
\label{sec::Basic}

In this section we are going to focus mainly on basic concepts and classical formulas to be able to compare them with our obtained results. First of all, we are going to study the properties of light-two level vapor interactions. Then we will generalize our results for multilevel systems in section \ref{sec::Selective}. The cell consists of three layers (glass, vapor, glass). Both boundaries are parallel to each other and light is directed normally to the first boundary. The constants (e.g.~dipole moment, resonance frequency) that we use here are taken from \cite{SteckRb852008}, where one can find physical parameters for rubidium atoms. The static magnetic field is set to zero and the cw laser field is considered to be weak to neglect all nonlinear effects associated with it.

We describe light-medium interactions by Maxwell equations using the density matrix formalism,

\begin{equation}
\begin{aligned}\label{eq::MaxwellEqWithDensity}
&\frac{d^2E}{dx^2} + k^2E = -4\pi k^2 P, \\
&\frac{d\rho}{dt} = - \frac{i}{\hbar} [\mathcal{H}\rho] + \Lambda,\\
&P = N Sp(d\rho),
\end{aligned}
\end{equation}

\noindent where $E$ is the amplitude of the electric field, $P$ is the polarization of the medium, N is the density of atoms, $d$ is the dipole moment of resonant atoms, $\mathcal{H}$ is the Hamiltonian of the system, and $\Lambda$ is the dissipation matrix, which describes all the relaxation processes, as well as the laser radiation linewidth. In the presence of spatial dispersion $P$ also is a function of the spatial coordinate (this case we will discuss in section \ref{sec::SpDisp}).

From the continuity of the electromagnetic field at the borders of the medium  ($x = 0$ and $x = L$) we have \cite{Dutier2003},

\begin{equation}
\label{eq::FieldCont_0}
\begin{aligned}
&E(0) = E_I + E_R,\\
&E'(0) = ikn_1(E_I - E_R),\\
&E(L) = E_T,\\
&E'(L) = ikn_2 E_T,
\end{aligned}
\end{equation}

\noindent where $E_I$ is the amplitude of the incident light, $E_R$ is the amplitude of the reflected light, $E_T$ is the amplitude of the transmitted light, $n_{1,2}$ are the refractive indexes of the windows, $k = \omega / c$ is the wave vector, $\omega$ is the frequency of the field and $c$ is the speed of light in vacuum.

In our calculations we will use the linear interaction approximation. This statement in density matrix formalism corresponds to $\rho_{11} \approx 1$ approximation. Namely, we assume that atoms are mainly in the ground state. For $\rho_{21}$ component of the density matrix we will have the following equation,

\begin{equation}
\label{eq::densityRho21}
u\frac{\partial \rho_{21}}{\partial x} = i \Omega e^{-i(\omega t - \varphi)} - (\Gamma + i\omega_0)\rho_{21},
\end{equation}

\noindent where  $\Omega = \frac{\abs{E}\abs{d}}{\hbar}$ is the Rabi frequency, $\Gamma = \gamma/2 + \Gamma_l + \Gamma_c + ...$ is the transverse decay rate, $\gamma$ is the natural decay rate of the excited state, $\Gamma_l$ is the laser spectral width, $\Gamma_c$ is a phenomenological decay rate that models the collisions, $u$ is the velocity of atoms, $\omega_0$ is the resonant frequency and $\varphi$ is the phase of the field.

For a very dense vapor, we can assume that the homogeneous width $\Gamma >> ku_T$ , where $ku_T$ is the Doppler width and $u_T$ is the most probable thermal velocity of atoms. Therefore, the $u\frac{\partial \rho_{21}}{\partial x}$ term in \eqref{eq::densityRho21} can be neglected \cite{Vartanyan1995}. So, from equations  (\ref{eq::MaxwellEqWithDensity}), (\ref{eq::densityRho21}) one can obtain classical formulas for susceptibility $\chi$ and refractive index $n$: 

\begin{equation}
\label{eq::RefAnsSucsConf}
\begin{gathered}
\chi = \frac{iq}{\Gamma-i\Delta},\\
n = \sqrt{1 + 4 \pi \chi},
\end{gathered}
\end{equation}

\noindent where $\Delta = \omega - \omega_0 $ is the detuning from the resonance frequency $\omega_0$, $q = \frac{N \abs{d}^2}{\hbar}$ is a parameter of the medium.

In order to take into account Doppler shift for a dilute gas in conventional theory one assumes that Doppler shift can be accounted for by simply replacing $\Delta$ in \eqref{eq::RefAnsSucsConf} by $\Delta - ku_T$ \cite{Vartanyan1995}. In this manner, we will obtain the steady-state solution. By using equations (\ref{eq::MaxwellEqWithDensity}), (\ref{eq::FieldCont_0}) and (\ref{eq::RefAnsSucsConf}) we can derive well known formulas for the field inside the medium and the reflection coefficient for Fabry-Perot interferometer,

\begin{equation}
\label{eq::LinSol}
E(x) = E_I A e^{-iknx} + E_I B e^{iknx},
\end{equation}

\noindent where 

\begin{equation}
\label{eq::LinSolCoef}
\begin{aligned}
&A = \frac{\widetilde{R} - r_1}{2n(n_1 + n)},\\
&B =  \frac{1 - \widetilde{R} r_1}{2n(n_1 + n)},\\
&\widetilde{R} = \frac{r_1 - r_2 e^{2iknL}}{1 - r_1 r_2 e^{2iknL}},\\
&r_{1,2} = \frac{n_{1,2} - n}{n_{1,2} + n},
\end{aligned}
\end{equation}

\noindent where $L$ is the length of the medium, $R = \abs{\widetilde{R}}$ is the reflection coefficient for the Fabry-Perot interferometer. If we set $n_1 = n_2 = n_0$ (and, therefore, $r_1 = r_2 = r$) we will obtain the following well-known classical formula for the reflection coefficient:

\begin{equation}
\label{eq::RClassical}
R = {\abs{\frac{r - r e^{2iknL}}{1 - r^2 e^{2iknL}}}}^2.
\end{equation}

The formulas presented in \eqref{eq::LinSolCoef} and \eqref{eq::RClassical} are well known from classical theory of the Fabry-Perot interferometer \cite{Svelto2009}. It is easy to show, that when $x \rightarrow L$ the first term in \eqref{eq::LinSol} is exponentially increasing ($Im(n) > 0$), while the second one is exponentially decreasing. However, the increasing term is not a problem. In order to show that we should represent the first term in the following way by substituting the expression of $\widetilde{R}$ into the expression of $A$ from \eqref{eq::LinSolCoef}:

\begin{equation}
\label{eq::LinSolSecondTerm}
E_I A e^{-iknx} \propto e^{2ikn(L-x/2)}.
\end{equation}

As can be seen from \eqref{eq::LinSolSecondTerm}, the first term of \eqref{eq::LinSol} is exponentially decreasing when $L \rightarrow \infty$. For long enough cell from \eqref{eq::LinSolCoef} we will have $\widetilde{R} = r_1$ and, consequently, $A = 0$. Therefore, for such cells we can neglect the first term.

For a more rigorous solution of \eqref{eq::densityRho21} one should not neglect $u\frac{\partial \rho_{21}}{\partial x}$ and find a solution that takes into account the transient behavior of the medium. This problem will be solved in the next section and we will compare our obtained formula for the reflection coefficient \eqref{eq::RDisp} with the classical one \eqref{eq::RClassical}. Also, we will show that the first term of our approximate solution \eqref{eq::EFieldInverseLaplaceWithR} has the same property as mentioned in \eqref{eq::LinSolSecondTerm}.

\section{Spatial Dispersion}
\label{sec::SpDisp}

Here we assume that the atoms are loosing their polarization after atom-wall collisions. In other words, atoms are leaving the wall in the ground state. This assumption can be described by boundary conditions for the medium polarization (see for example \cite{Dutier2003, Vartanyan1995}):

\begin{equation}\label{eq::PolBoundery}
P(x = 0, u > 0) = 0,
\qquad
P(x = L, u < 0) = 0.
\end{equation}

The effect of loosing the polarization gives rise to the spatial dispersion. That is to say, the susceptibility depends also on the spatial coordinate. From equations (\ref{eq::MaxwellEqWithDensity}) and (\ref{eq::PolBoundery}) one can derive

\begin{equation}\label{eq::SpatialPolariztion}
P(x) = \int_{0}^{x} E(y) \langle \chi(x-y) \rangle_{u>0} dy   + \int_{L}^{x} E(y) \langle \chi(x-y) \rangle_{u<0} dy,
\end{equation}

\noindent where $\chi(x) = \frac{iq}{u} e^{-\frac{\Gamma - i\Delta}{u} x}$
is the linear susceptibility for the medium with a spatial dispersion and $\langle ... \rangle_{u}$ denotes averaging over velocities (Here we assume Maxwellian distribution). 

It's hard to solve the differential equation in \eqref{eq::MaxwellEqWithDensity} with the polarization from \eqref{eq::SpatialPolariztion}. First of all let's find the asymptotic solution when $x \rightarrow L$. In this solution we will neglect the second integral, because it vanishes, when $x \rightarrow L$. The resulting differential equation will be

\begin{equation}\label{eq::MaxwellEqOnly}
\frac{d^2E(x)}{dx^2} + k^2E(x) = -4\pi k^2 \int_{0}^{x} E(y) \langle \chi(x-y) \rangle_{u>0} dy.
\end{equation}

After a Laplace transformation we will obtain

\begin{equation}\label{eq::LaplaceEField}
\widetilde{E}(s) = \frac{sE(0) + E'(0)}{s^2 + k^2(1 + 4 \pi \widetilde{\chi}(s)_{u > 0})},
\end{equation}

\noindent where $\widetilde{\chi}$ is the Laplace transform of $\langle \chi(x-y) \rangle_{u>0}$, $\widetilde{E}(s)$ is the Laplace transform of $E(x)$ and $s$ is the Laplace variable. For two level system the explicit form of $\widetilde{\chi}$ is

\begin{equation}
\label{eq::SuscepSpatialLaplace}
\widetilde{\chi}(s)= {\langle \frac{i q}{\Gamma + s u - i\Delta} \rangle_{u>0}}.
\end{equation}

\begin{figure}[h!]
	\begin{center}
		\includegraphics[width = 240pt, height=240pt]{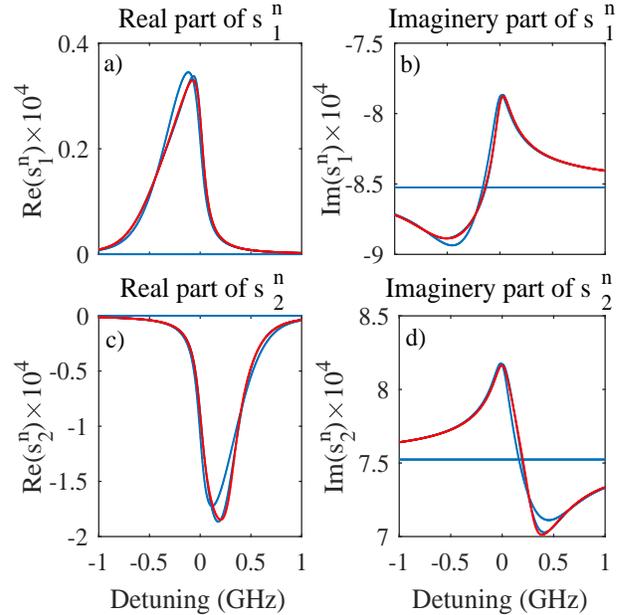}
		\caption{\label{fig:IntProc} Convergence of the iteration procedure. a) and b) correspond to real and imaginary parts of $s_1$ and c) and d) correspond to real and imaginary parts of $s_2$. In all four plots the red solid lines correspond to 10th iteration from \eqref{eq::intproc}, blue solid horizontal lines correspond to $n=0$ iteration step,  and the rest blue solid lines correspond to $n=1,2$ iterations. The density of atoms is $N = 10^{15}cm^{-3}$ and $\Gamma_{c} = 2 \pi \cdot  53 MHz$.}
	\end{center}
\end{figure}

Here we will do the following approximation. We will represent denominator of \eqref{eq::LaplaceEField} in this way 

\begin{equation}\label{eq::DispEq}
s^2 + k^2(1 + 4 \pi \widetilde{\chi}(s)_{u > 0}) = (s - s_1)(s - s_2).
\end{equation}

\noindent where $s_1$ and $s_2$ are given by the following iteration procedure:

\begin{equation}
\label{eq::intproc}
\begin{aligned}
&s^{(n)}_1 = -i k (1 + 2 \pi \widetilde{\chi}(s^{n-1}_1)) = -ik n(s^{n-1}_1),\\
&s^{(n)}_2 = i k (1 + 2 \pi \widetilde{\chi}(s^{n-1}_2))) = ik n(s^{n-1}_2),\\
&s^{(0)}_1 = - ik,
\qquad s^{(0)}_2 =  ik,
\end{aligned}
\end{equation}

\noindent here $n$ is the iteration step. With each iteration step we will obtain more precise values for $s_1$ and $s_2$. This iteration procedure works when $ 4 \pi \abs{\widetilde{\chi}(s_{1,2})_{u > 0}} \ll 1$. Note that $s^{(0)}_1$ and $s^{(0)}_2$ are the roots of the denominator of \eqref{eq::LaplaceEField}, when $\chi(s)_{u > 0} = 0$. If $4 \pi \abs{\chi(s_{1,2})_{u > 0}} \ll 1$, the iteration procedure will give the approximate roots of the denominator of \eqref{eq::LaplaceEField}. Convergence of the iteration procedure is demonstrated in Fig.\ref{fig:IntProc}.

\begin{figure}[h!]
	\begin{center}
		\includegraphics[width = 250pt, height=280pt]{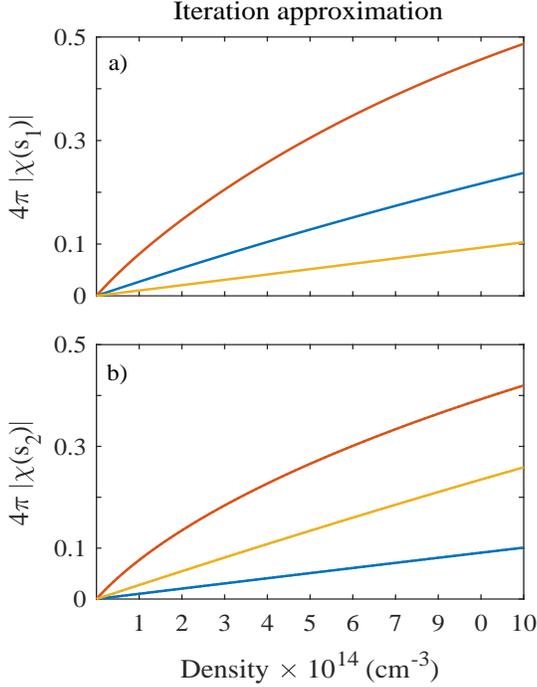}
		\caption{\label{fig:IntAprox} Dependence of $4 \pi \widetilde{\chi}(s_{k}))_{u > 0}$ (where $k=1,2$ corresponds respectfully to sub-figures a) and b) on the density of atoms. The red lines correspond to $\Delta = 0$, the yellow lines to $\Delta = 0.5GHz$ and the blue ones to $\Delta = -0.5GHz$.}
	\end{center}
\end{figure}

From Fig.\ref{fig:IntProc} one can see that only three iterations are needed in order to find $s_1$ and $s_2$. We should note that for densities $N>10^{15}cm^{-3}$ the iteration procedure doesn't work, because $ 4 \pi \abs{\widetilde{\chi}(s_{1,2})_{u > 0}} \ll 1$ is not true for these densities. The dependence of $4 \pi \abs{\widetilde{\chi}(s_{1,2}))_{u > 0}}$ on the density of atoms is shown in Fig.\ref{fig:IntAprox}.

So, by taking into account \eqref{eq::DispEq} we can rewrite \eqref{eq::LaplaceEField} as follows:

\begin{equation}\label{eq::LaplaceEFieldRoots}
\widetilde{E}(s) = \frac{sE(0) + E'(0)}{(s - s_1)(s - s_2)}.
\end{equation}

After an inverse Laplace transformation from \eqref{eq::LaplaceEFieldRoots} one can obtain an expression for $E(x)$,

\begin{equation}
\label{eq::EFieldInverseLaplace}
E(x) =  \frac{s_1 E(0) + E'(0)}{s_1 - s_2} e^{s_1 x} - \frac{s_2 E(0) + E'(0)}{s_1 - s_2} e^{s_2 x}.
\end{equation}

Here we should remember that \eqref{eq::EFieldInverseLaplace} is an approximative solution, hence, it is correct only for $x \rightarrow L$. With the use of \eqref{eq::EFieldInverseLaplace} we can find $E(L)$ and E'(L):

\begin{equation}
\label{eq::ELAndE'L}
\begin{aligned}
&E(L) =  \frac{s_1 E(0) + E'(0)}{s_1 - s_2} e^{s_1 L} - \frac{s_2 E(0) + E'(0)}{s_1 - s_2} e^{s_2 L},\\
&E'(L) = \frac{{s_1}^2 E(0) + s_1 E'(0)}{s_1 - s_2} e^{s_1 L} - \frac{{s_2}^2 E(0) + s_2 E'(0)}{s_1 - s_2} e^{s_2 L}.
\end{aligned}
\end{equation}

From \eqref{eq::ELAndE'L} together with the conditions from \eqref{eq::FieldCont_0} one can obtain an expression for $\widetilde{R} = E_R / E_I$:

\begin{equation}
\label{eq::ReflectionDisp}
\begin{aligned}
&\widetilde{R} = \frac{r_1(s_1) - D  r_2(s_2) e^{\phi}}{1 - D  r_1(s_2) r_2(s_2) e^{\phi}},\\
&D = \frac{(n_1 + n(s_2))(n_2 + n(s_2)}{(n_1 + n(s_1))(n_2 + n(s_1))},\\
&r_l(s_m) = \frac{n_l - n(s_m)}{n_l + n(s_m)},\\
&\phi = 2 ik n_{avg} L,
\end{aligned}
\end{equation}

\noindent where $l,m = 1,2$ are integers and $n_{avg} = (n(s_1) + n(s_2)) / 2$.

In \eqref{eq::EFieldInverseLaplace} one can notice two exponential expressions, one of which is increasing ($Re(s_1) > 0$) and the other one is decreasing ($Re(s_2) < 0$), when $x \rightarrow \infty$. So, a natural question arises: when $L \rightarrow \infty$ and, consequently, $x$ can increase to infinity, can the first term become infinite? If so, this solution will be unphysical (the field should be zero at infinity). To show that there is no problem with the first term, we will rewrite \eqref{eq::EFieldInverseLaplace} by using the conditions from \eqref{eq::FieldCont_0} in the following way:

\begin{equation}
\label{eq::EFieldInverseLaplaceWithR}
\begin{aligned}
&E(x) = E_I A e^{s_1 x} + E_I B e^{s_2 x}, \\
&A = \frac{\widetilde{R} - r_1(s_1)}{(n(s_1) + n(s_2))(n_1 + n(s_1))}, \\
&B = \frac{1 - \widetilde{R} r_1(s_2)}{(n(s_1) + n(s_2))(n_1 + n(s_2))}. \\
\end{aligned}
\end{equation}

By substituting the expression of $\widetilde{R}$ from \eqref{eq::ReflectionDisp} to the first term of \eqref{eq::EFieldInverseLaplaceWithR} we will obtain

\begin{equation}
\label{eq::EFieldSecondTerm_Disp}
E_I A e^{s_1 x} \propto e^{s_2 L - s_1(L-x)}.
\end{equation}

\noindent From \eqref{eq::EFieldSecondTerm_Disp} one can notice that when $L \rightarrow \infty$ that expression tends to zero for every $x \in (0, L)$. So, when we have only the first border (or the second border is far enough) we can neglect the exponentially increasing term in \eqref{eq::EFieldInverseLaplaceWithR}. Therefore, here, like in \eqref{eq::LinSol}, there are no problems with infinities.

If we set $n_1 = n_2 = n_0$ in the expression for $\widetilde{R}$ from \eqref{eq::ReflectionDisp} we can obtain a more simpler expression for the reflection coefficient $R = \abs{{\widetilde{R}}}^2$:

\begin{equation}
\label{eq::RDisp}
R = {\abs{\frac{r(s_1) - D r(s_2) e^\phi}{1 - D r^2(s_2) e^\phi}}}^2.
\end{equation}

In section \ref{sec::RefractiveIndex} this result will be compared with the classical formula from \eqref{eq::RClassical}. Also, in the same section we will discuss physical meaning of $n(s_1)$ and $n_{avg}$.

\section{Effective refractive index}
\label{sec::RefractiveIndex}

Diffusive atom collisions with the wall create a transient spatial region near the boundaries of the cell. In this transient region we do not have one uniform vapor, thus, we can not describe our medium with the linear refractive index \cite{Schuurmans1980}. From \eqref{eq::SpatialPolariztion} one observes nonlocal dependence of the polarization on the spatial coordinate $x$. Thus, the refractive index should have dependence on $x$. Although, it is hard to derive the expression of the refractive index for this kind of medium, we are able to attribute the concept of effective refractive index to the medium. 

\begin{figure}[h!]
	\begin{center}
		\includegraphics[width = 250pt, height=250pt]{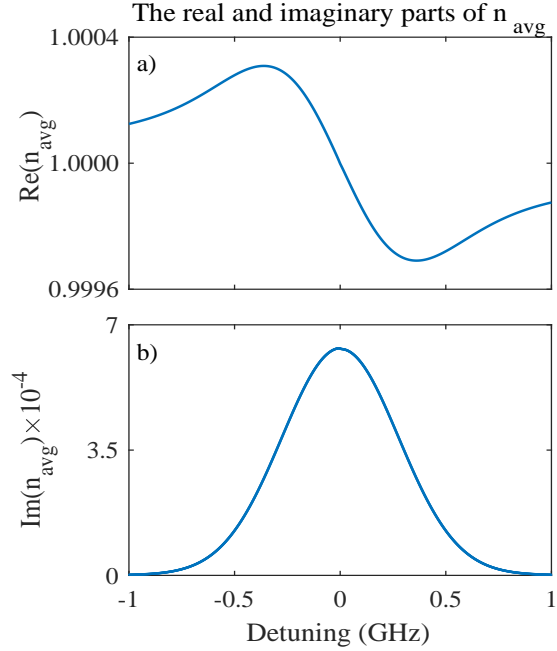}
		\caption{\label{fig:ReImN_avg} The spectra of a) real and b) imaginary parts of $n_{avg}$. The density of atoms is $N = 4 \cdot 10^{12}cm^{-3}$.}
	\end{center}
\end{figure}

First of all, let us compare our obtained formula from \eqref{eq::RDisp} with the classical formula for the reflection coefficient presented in \eqref{eq::RClassical}. In our formula, instead of $r$ we have $r(s_1)$ and $r(s_2)$. Also, we have an additional term  $D$. Actually $D$ like all terms of \eqref{eq::RDisp} has a dependence on the detuning $\Delta$, but it is always close to one and can be neglected. Finally, notice that the refractive index $n$ in \eqref{eq::RClassical} and $n_{avg}$ from \eqref{eq::RDisp} are both in the phases of exponents in the corresponding formulas. So, $n_{avg}$ in \eqref{eq::RDisp} plays the same role as $n$ in \eqref{eq::RClassical}. This comparison can lead to an assumption that $n_{avg}$ can play the role of the refractive index in our medium, but it can not be regarded as actual refractive index, because it doesn't depend on the spatial coordinate (the refractive index of the medium with the spatial dispersion should have a dependence on $x$). The real and imaginary parts of $n_{avg}$ are presented in Fig.\ref{fig:ReImN_avg}.

As you can see from Fig.\ref{fig:ReImN_avg} the curve of the real part of $n_{avg}$ has a dispersive profile and the curve of the imaginary part of $n_{avg}$ has a absorptive profile, like the refractive index in the conventional theory \cite{Landsberg2003}. This is another argument that $n_{avg}$ has a physical meaning of the refractive index inside the medium.

From \eqref{eq::RDisp} we can derive another interesting formula, if we assume that the cell is long enough ($L\rightarrow \infty$). Notice that $e^{\phi} \rightarrow 0$, when $L \rightarrow \infty$ ($Im(n_{avg}) > 0$). Hence, we will obtain the following formula:

\begin{equation}
\label{eq::SingleRef}
R_s = \abs{r(s_1)}^2 = \abs{\frac{n_0 - n(s_1)}{n_0 + n(s_1)}}^2.
\end{equation}

\begin{figure}[h!]
	\begin{center}
		\includegraphics[width = 250pt, height=250pt]{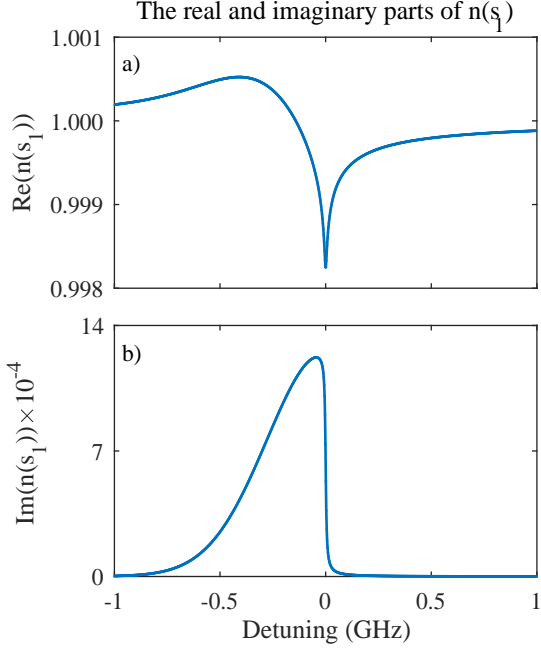}
		\caption{\label{fig:DispAbs} The spectra for a) the effective real refractive index and b) the effective absorption. The density of atoms is $N = 4 \cdot 10^{12}cm^{-3}$.}
	\end{center}
\end{figure}

In this formula only the first reflection from the cell is taken into account, thus, it represents the formula for the single selective reflection. Notice that \eqref{eq::SingleRef} is similar to the Fresnel equation for normal incidence. Therefore, we can say that $n(s_1)$ plays the role of the refractive index. So, we will call $n(s_1)$ the effective complex refractive index for a single reflection. Also, by analogy to the Fresnel equation, we will attribute $Im(n(s_1))$ as the effective absorption, and $Re(n(s_1))$ as the effective real refractive index. Moreover, $n(s_1)$ can also be referred to as the surface admittance $M = E'(0)/(ikE(0))$ as defined in \cite{Schuurmans1976}. In Fig.\ref{fig:DispAbs} we show the dependence of $Im(n(s_1))$ and $Re(n(s_1))$ on the detuning from the resonance line.

In the effective absorption curve in Fig.\ref{fig:DispAbs}$b$ one can see that the absorption curve is red-shifted. This means that the atoms with positive velocities (their Doppler shift is red-shifted) have dominant contribution in the  absorption curve. It is interesting to recall the experimental results obtained by Burgmans et al. \cite{Burgmans1977}, where the authors observed fluorescence radiation from the transient region of the Na vapor. They showed that the spectrum of fluorescence of atoms near the wall has a decrease in the blue-shifted sides of resonance lines. The reason is the possibility for the atoms that have a polarization and are moving towards the boundary to lose their polarization non-radiatively by quenching to the wall. In Fig.\ref{fig:DispAbs} one can notice that the effective absorption curve is different from the absorption curve described by the conventional theory \cite{Landsberg2003}. The difference, like in \cite{Burgmans1977} experiment, is in the blue-shifted side from the resonance line. So, in this sense, we can claim that our result is consistent with that experiment.

\section{Selective reflection}
\label{sec::Selective}

\begin{figure}[h!]
	\begin{center}
		\includegraphics[width = 250pt, height=150pt]{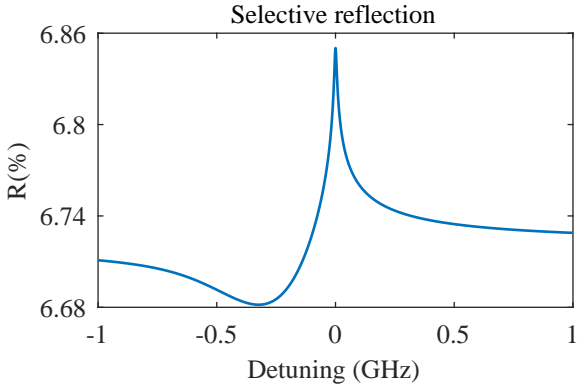}
		\caption{\label{fig:SingleSR} The selective reflection spectrum in the vicinity of the resonance line. The density of atoms is $N = 10^{13}cm^{-3}$.}
	\end{center}
\end{figure}

In Fig.\ref{fig:SingleSR} we show the dispersion of the single selective reflection calculated from \eqref{eq::SingleRef}.

\begin{figure}[h!]
	\begin{center}
		\includegraphics[width = 250pt, height=150pt]{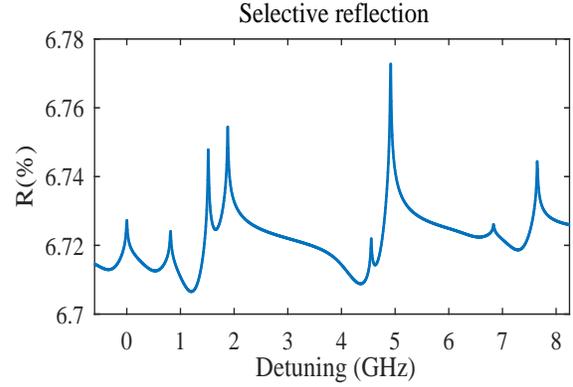}
		\caption{\label{fig:RbD1SR} The selective reflection spectra from all $\mathbf{Rb}$'s D$_1$ lines (including both $\prescript{85}{}{\mathbf{Rb}}$ and $\prescript{87}{}{\mathbf{Rb}}$). The density of atoms is $N = 10^{13} cm^{-3}$. }
	\end{center}
\end{figure}

The result presented in Fig.\ref{fig:SingleSR} is similar to the well known profile of the selective reflection presented, for example, in figure 8 from \cite{Bloch2005}. Of course, it is a theoretical simplification to assume that we have only a single resonance line. In real experiments it is useful to consider multilevel systems. To generalize our theory to the multilevel system we need to make an assumption that the light-medium interaction is linear. In this case we should leave the same all the calculations mentioned above, but we need to change the expression for $\widetilde{\chi}(s)$ presented in \eqref{eq::SuscepSpatialLaplace}. So, for example, the susceptibility of $D_1$ line of rubidium vapor will be as following:

\begin{equation}
\label{eq::suscepD1}
\widetilde{\chi}(s)=\sum_{k=1}^{8} {\langle \frac{iq_k}{su + \Gamma_k - i\Delta_k} \rangle_{u>0}},
\end{equation}

\noindent where $q_k = N \abs{d_k}^2 / \hbar$ is a parameter, $d_k$ are the dipole moments, $\Delta_k = \omega - \omega_k$ are the detunings from the corresponding resonance frequencies $\omega_k$,  $\Gamma_k$ are the homogeneous widths of the corresponding F$\rightarrow$F' hyperfine transitions for rubidium atomic vapor. The calculated spectrum of the rubidium D$_1$ line is shown in Fig.\ref{fig:RbD1SR}.

In Fig.\ref{fig:RbD1SR} we assume natural abundance for rubidium atoms ($72.2\% \prescript{85}{}{\mathbf{Rb}}$ and $27.8\% \prescript{87}{}{\mathbf{Rb}}$). Here we also should mention that in \eqref{eq::suscepD1} we don't take into account the depopulation of levels by assuming that it has a small effect on our model. The result can be compared to the experimental results obtained by Wang et al. \cite{Wang1997} and Badalyan et al. \cite{Badalyan2006} presented in the corresponding figures for rubidium D$1$ lines. Our result is consistent with the experimental curves presented there, although in these two papers the order of the density of the vapor is $N \approx 10^{14} cm^{-3}$.

\begin{figure}[h!]
	\begin{center}
		\includegraphics[width = 250pt, height=150pt]{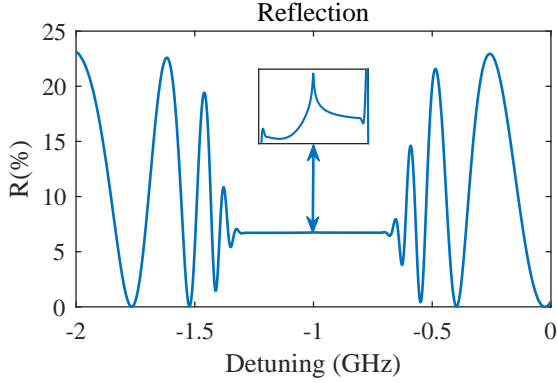}
		\caption{\label{fig:RFabry} The multiple reflection spectrum from the Fabry-Perot interferometer in the far wings of the resonance line. In the insertion we show the zoomed image of the horizontal line from the reflection spectrum. The density of atoms is $N = 3 \cdot 10^{13} cm^{-3}$. }
	\end{center}
\end{figure}

Another interesting spectrum can be obtained by increasing the scale of detuning for the selective reflection. In this case we can see from Fig.\ref{fig:RFabry} that we have oscillations in the "wings" of the resonance line.

In Fig.\ref{fig:RFabry} the selective reflection profile presented in Fig.\ref{fig:SingleSR} also exists. Selective reflection profile is "hidden" in the region where we have a horizontal line in Fig.\ref{fig:RFabry}. To see this one should zoom in this region and a picture like in Fig.\ref{fig:SingleSR} will emerge (see the zoomed image). Fig.\ref{fig:RFabry} is interesting, because from this spectrum it is easy to straightforwardly obtain the group refractive index as presented in \cite{Papoyan2017}.

\begin{figure}[h!]
	\begin{center}
		\includegraphics[width = 250pt, height=250pt]{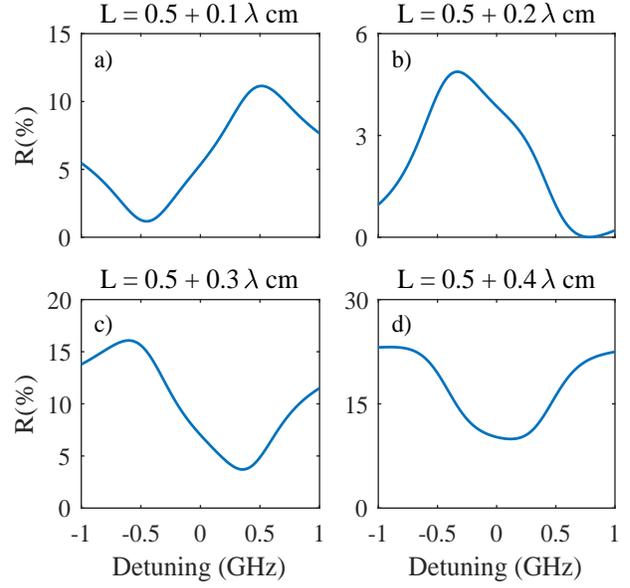}
		\caption{\label{fig:4Length} The selective reflection spectrum for different cell lengths. The subfigures a-d are the spectra from the cells with length varying from $L \approx 0.5 + 0.1\lambda\,cm$ to $L \approx 0.5 + 0.4\lambda\,cm$. The density of atoms is $N = 10^{11} cm^{-3}$. }
	\end{center}
\end{figure}

In regions near the resonance in Figs.\ref{fig:SingleSR}, \ref{fig:RbD1SR} and in the horizontal region of Fig.\ref{fig:RFabry} we have only a single reflection from the first boundary. For the multiple reflection spectrum the light should be able to reach the second boundary, to reflect from it and, finally, to be able to reach and to pass the first boundary of the cell. If this doesn't happen, because the absorption is high in the vicinity of the resonance line, there will be only a single reflection. To see a multiple reflection one should either decrease the length of the cell, or decrease the density of the vapor. In Fig.\ref{fig:4Length} we show the selective reflection from the cell with length $L \approx 0.5cm$ and density $N = 10^{11} cm^{-3}$, calculated from \eqref{eq::RDisp}.

In Fig.\ref{fig:4Length} one can see different profiles of the selective reflection that correspond to different lengths of the medium. The differences arise from the fact that reflection is not generated from a single reflection, but from multiple reflections from the cell. Consequently, we will have interference picture from all those reflections. When we vary the length of the cell the interference pattern changes and the profile of reflection also changes by its shape and amplitude. We should note here that after adding $0.5\lambda$ to $L$ the same picture will be obtained. So, the pattern repeats itself after every $0.5\lambda$. This repeating pattern is well known from the literature \cite{Svelto2009}.

\section{Conclusion}
\label{sec::Conclusion}

In this paper we developed a self-consistent theory for the selective reflection from a dilute vapor. We obtained formulas for single and multiple reflections. We obtained the spectrum for the effective refractive index that is also known as the surface admittance \cite{Schuurmans1976}. We can say that our results are well consistent with the above mentioned experimental results. Here we should notice that our theory is developed for a dilute vapor. When the medium is dense we should take into account effects like radiation trapping, the Dicke narrowing \cite{Dicke1953} and etc. Also, in our model we did not take into account various shifts that are present in the selective reflection spectrum (e.g.~caused by local-field correction, atom-wall
interaction, non-exponential attenuation of the field in the vapor \cite{guo1996}).

Our developed model can be applied in a wide range of physical problems. For example, it can be used for determination of the density of unwanted (or in other cases desired) vapors by the selective reflection spectrum. If unwanted atoms have high enough density, the signal from Fabry-Perot interferometer will be a single selective reflection that can be compared with the spectrum calculated from \eqref{eq::SingleRef} and used to find the density of unwanted atoms. In the case when Fabry-Perot interferometer has a small length and the density of atoms is low enough that there will be multiple reflections from the boundaries of the cell, \eqref{eq::RDisp} can be used. Also, with this technique one can change the spectral profile of radiation by manipulating the length and the density of the dilute vapor cell, as presented in Fig.\ref{fig:4Length}. This theory can be generalized for multiphoton interactions that will provide an opportunity to address problems like, for example, selective reflection in EIT configuration \cite{Fleischhauer2005}.

\section{Acknowledgments}
\label{sec::Acknowledgments}

The author is grateful to Gayane Grigoryan for stimulating discussions and valuable advices.


\begin{thebibliography}{10}
	\newcommand{\enquote}[1]{``#1.''}
	
	\bibitem{woerdman1975}
	J.~Woerdman and M.~Schuurmans, \enquote{Spectral narrowing of selective
		reflection from sodium vapour,} {{Optics
			Communications}} \textbf{14}, 248 -- 251 (1975).

	\bibitem{Schuurmans1976}
	M.~F.~H. Schuurmans, \enquote{Spectral narrowing of selective reflection,}
	{{J. Phys. France}} \textbf{37}, 469--485 (1976).
	
	\bibitem{gazazyan2007}
	E.~A. Gazazyan, A.~V. Papoyan, D.~Sarkisyan, and A.~Weis, \enquote{Laser
		frequency stabilization using selective reflection from a vapor cell with a
		half-wavelength thickness,} {{Laser Physics Letters}}
	\textbf{4}, 801 (2007).
	
	\bibitem{li1998}
	R.~Li, S.~Jia, D.~Bloch, and M.~Ducloy, \enquote{Frequency-stabilization of a
		diode laser with ultra-low power through linear selective reflection,}
	{{Optics Communications}} \textbf{146}, 186 -- 188
	(1998).
	
	\bibitem{muller1998}
	R.~M{\"u}ller and A.~Weis, \enquote{Laser frequency stabilization using
		selective reflection spectroscopy,} {{Applied Physics
			B}} \textbf{66}, 323--326 (1998).
	
	\bibitem{Papoyan2017}
	A.~Papoyan, S.~Shmavonyan, D.~Khachatryan, and G.~Grigoryan,
	\enquote{Straightforward retrieval of dispersion in a dense atomic vapor
		helped by buffer gas-assisted radiation channeling,}
	{{J. Opt. Soc. Am. B}} \textbf{34}, 877--883 (2017).
	
	\bibitem{Klinger2017JETP}
	E.~Klinger, A.~Sargsyan, C.~Leroy, and D.~Sarkisyan, \enquote{Selective
		reflection of laser radiation from submicron layers of rb and cs atomic
		vapors: Applications in atomic spectroscopy,} {{Journal
			of Experimental and Theoretical Physics}} \textbf{125}, 543--550 (2017).
	
	\bibitem{Sargsyan2016}
	A.~Sargsyan, E.~Klinger, Y.~Pashayan-Leroy, C.~Leroy, A.~Papoyan, and
	D.~Sarkisyan, \enquote{Selective reflection from rb vapor in half- and
		quarter-wave cells: Features and possible applications,}
	{{JETP Letters}} \textbf{104}, 224--230 (2016).
	
	\bibitem{failache2003}
	H.~Failache, S.~Saltiel, M.~Fichet, D.~Bloch, and M.~Ducloy, \enquote{Resonant
		coupling in the van der waals interaction between an excited alkali atom and
		a dielectric surface: an experimental study via stepwise selective reflection
		spectroscopy,} {{The European Physical Journal D -
			Atomic, Molecular, Optical and Plasma Physics}} \textbf{23}, 237--255 (2003).
	
	\bibitem{fichet2007}
	M.~Fichet, G.~Dutier, A.~Yarovitsky, P.~Todorov, I.~Hamdi, I.~Maurin,
	S.~Saltiel, D.~Sarkisyan, M.-P. Gorza, D.~Bloch, and M.~Ducloy,
	\enquote{Exploring the van der waals atom-surface attraction in the
		nanometric range,} {{EPL (Europhysics Letters)}}
	\textbf{77}, 54001 (2007).
	
	\bibitem{Sargsyan2017OpLe}
	A.~Sargsyan, A.~Papoyan, I.~G. Hughes, C.~S. Adams, and D.~Sarkisyan,
	\enquote{Selective reflection from an rb layer with a thickness below
		$\lambda$/12 and applications,} {{Opt. Lett.}}
	\textbf{42}, 1476--1479 (2017).
	
	\bibitem{vuletic1993}
	V.~Vuletić, V.~Sautenkov, C.~Zimmermann, and T.~Hänsch, \enquote{Measurement
		of cesium resonance line self-broadening and shift with doppler-free
		selective reflection spectroscopy,} {{Optics
			Communications}} \textbf{99}, 185 -- 190 (1993).
	
	\bibitem{Wang1997}
	P.~Wang, A.~Gallagher, and J.~Cooper, \enquote{Selective reflection by rb,}
	{{Phys. Rev. A}} \textbf{56}, 1598--1606 (1997).
	
	\bibitem{papoyan1998Izvest}
	A.~Papoyan, \enquote{Measurement of collisional self broadening of atomic
		resonance lines in selective reflection experiment,}
	{{J. Contemp. Phys. (Arm. Acad. Sci.)}} \textbf{33},
	109--114 (1998).
	
	\bibitem{papoyan1998OpSp}
	A.~Papoyan, G.~Sarkisyan, and S.~Shmavonyan, \enquote{Selective reflection of
		light from dense sodium vapors,} {{Optics and
			Spectroscopy}} \textbf{85}, 649--652 (1998).
	
	\bibitem{gross1997}
	B.~Gross, N.~Papageorgiou, V.~Sautenkov, and A.~Weis, \enquote{Velocity
		selective optical pumping and dark resonances in selective reflection
		spectroscopy,} {{Phys. Rev. A}} \textbf{55}, 2973--2981
	(1997).
	
	\bibitem{weis1992}
	A.~Weis, V.~A. Sautenkov, and T.~W. H\"ansch, \enquote{Observation of
		ground-state zeeman coherences in the selective reflection from cesium
		vapor,} {{Phys. Rev. A}} \textbf{45}, 7991--7996 (1992).
	
	\bibitem{papageorgiou1994}
	N.~Papageorgiou, A.~Weis, V.~A. Sautenkov, D.~Bloch, and M.~Ducloy,
	\enquote{High-resolution selective reflection spectroscopy in intermediate
		magnetic fields,} {{Applied Physics B}} \textbf{59},
	123--126 (1994).
	
	\bibitem{Sargsyan2017JOSAB}
	A.~Sargsyan, E.~Klinger, G.~Hakhumyan, A.~Tonoyan, A.~Papoyan, C.~Leroy, and
	D.~Sarkisyan, \enquote{Decoupling of hyperfine structure of cs d1 line in
		strong magnetic field studied by selective reflection from a nanocell,}
	{{J. Opt. Soc. Am. B}} \textbf{34}, 776--784 (2017).
	
	\bibitem{Klinger2017Euro}
	E.~Klinger, A.~Sargsyan, A.~Tonoyan, G.~Hakhumyan, A.~Papoyan, C.~Leroy, and
	D.~Sarkisyan, \enquote{Magnetic field-induced modification of selection rules
		for rb d2 line monitored by selective reflection from a vapor nanocell,}
	{{The European Physical Journal D}} \textbf{71}, 216
	(2017).
	
	\bibitem{Sargsyan2018}
	A.~Sargsyan, A.~Tonoyan, J.~Keaveney, I.~G. Hughes, C.~S. Adams, and
	D.~Sarkisyan, \enquote{Selective reflection of potassium vapor nanolayers in
		a magnetic field,} {{Journal of Experimental and
			Theoretical Physics}} \textbf{126}, 293--301 (2018).
	
	\bibitem{ducloy1991}
	{M. Ducloy} and {M. Fichet}, \enquote{General theory of frequency modulated
		selective reflection. influence of atom surface interactions,}
	{{J. Phys. II France}} \textbf{1}, 1429--1446 (1991).
	
	\bibitem{Vartanyan2001}
	T.~A. Vartanyan and A.~Weis, \enquote{Origin of the ``blueshift'' in selective
		reflection spectroscopy and its partial compensation by the local-field
		correction,} {{Phys. Rev. A}} \textbf{63}, 063813
	(2001).
	
	\bibitem{guo1994}
	J.~Guo, J.~Cooper, A.~Gallagher, and M.~Lewenstein, \enquote{Theory of
		selective reflection spectroscopy,} {{Optics
			Communications}} \textbf{110}, 732 -- 743 (1994).
	
	\bibitem{guo1996}
	J.~Guo, J.~Cooper, and A.~Gallagher, \enquote{Selective reflection from a dense
		atomic vapor,} {{Phys. Rev. A}} \textbf{53}, 1130--1138
	(1996).
	
	\bibitem{Vartanyan1995}
	T.~A. Vartanyan and D.~L. Lin, \enquote{Enhanced selective reflection from a
		thin layer of a dilute gaseous medium,} {{Phys. Rev. A}}
	\textbf{51}, 1959--1964 (1995).
	
	\bibitem{Dutier2003}
	G.~Dutier, S.~Saltiel, D.~Bloch, and M.~Ducloy, \enquote{Revisiting optical
		spectroscopy in a thin vapor cell: mixing of reflection and transmission as a
		fabry--perot microcavity effect,} {{J. Opt. Soc. Am. B}}
	\textbf{20}, 793--800 (2003).
	
	\bibitem{Burgmans1977}
	A.~L.~J. Burgmans, M.~F.~H. Schuurmans, and B.~B\"olger, \enquote{Transient
		behavior of optically excited vapor atoms near a solid interface as observed
		in evanescent wave emission,} {{Phys. Rev. A}}
	\textbf{16}, 2002--2007 (1977).
	\newpage		
	\bibitem{bordo1997}
	V.~Bordo, C.~Henkel, A.~Lindinger, and H.-G. Rubahn, \enquote{Evanescent wave
		fluorescence spectra of na atoms,} {{Optics
			Communications}} \textbf{137}, 249 -- 253 (1997).

	\bibitem{Papoyan2004}
	A.~V. Papoyan, G.~G. Grigoryan, S.~V. Shmavonyan, D.~Sarkisyan, J.~Gu{\'e}na,
	M.~Lintz, and M.~A. Bouchiat, \enquote{New feature in selective reflection
		with a highly parallel window: phase-tunable homodyne detection of the
		radiated atomic field,} {{The European Physical Journal
			D - Atomic, Molecular, Optical and Plasma Physics}} \textbf{30}, 265--273
	(2004).
	
	\bibitem{SteckRb852008}
	D.~A. Steck, \enquote{Rubidium 85 d line data,}  (2008).
	
	\bibitem{Svelto2009}
	O.~Svelto, \emph{Principles of Lasers} (Springer US, 2010), 5th ed.

	\bibitem{Schuurmans1980}
	M.~H. Schuurmans, \enquote{The fluorescence of atoms near a glass surface,}
	{{Contemporary Physics}} \textbf{21}, 463--482 (1980).

	\bibitem{Landsberg2003}
	G.~S. Landsberg, \emph{Optika (Optics)} (Fizmatlit, 2003), 6th ed.
	
	\bibitem{Bloch2005}
	D.~Bloch and M.~Ducloy, \enquote{Atom-wall interaction,}  (Academic Press,
	2005), pp. 91 -- 154.
	
	\bibitem{Badalyan2006}
	A.~Badalyan, V.~Chaltykyan, G.~Grigoryan, A.~Papoyan, S.~Shmavonyan, and
	M.~Movsessian, \enquote{Selective reflection by atomic vapor: experiments and
		self-consistent theory,} {{The European Physical Journal
			D - Atomic, Molecular, Optical and Plasma Physics}} \textbf{37}, 157--162
	(2006).
	
	\bibitem{Dicke1953}
	R.~H. Dicke, \enquote{The effect of collisions upon the doppler width of
		spectral lines,} {Phys. Rev.} \textbf{89}, 472--473
	(1953).
	
	\bibitem{Fleischhauer2005}
	M.~Fleischhauer, A.~Imamoglu, and J.~P. Marangos, \enquote{Electromagnetically
		induced transparency: Optics in coherent media,} {{Rev.
			Mod. Phys.}} \textbf{77}, 633--673 (2005).
	
\end{thebibliography}
\end{document}